\documentclass[11pt,preprint]{elsarticle}

\usepackage{bbm,enumerate}
\usepackage{graphicx,pstricks}
\usepackage{pst-plot}
\usepackage{pst-tree}
\usepackage{pst-node}
\usepackage{amsfonts,amsmath,amssymb}
\usepackage{epsfig}	
\usepackage{colordvi}
\usepackage{color}

\newcommand{\remove}[1]{}

\newtheorem{theorem}{Theorem}
\newtheorem{lemma}{Lemma}

\newtheorem{definition}{Definition}

\newtheorem{example}{Example}

\def\proof{\noindent{\bf Proof. $\,$}}

\def\influenced{{\tt Influenced}}
\def\active{{\tt Active}}

\def\PROB{{TWC--TSS }}
\def\minrep{{\tt MIN$\,$REP }}

\def\Ch{{\tt Ch}}
\def\SS{{ S}}
\def\SN{{\tt s}}

\def\PREC{{\tt J}}

\def\r{{\bf r}}
\def\k{{ d}}
\def\i{{ r}}
\def\cA{{\cal A}}
\def\cB{{\cal B}}
\def\span{{time  window }}
\def\diameter{{diam }}
\def\root{{p }}
\newcommand{\N}{{\mathbb{N}}}

\journal{Theoretical Computer Science}

\begin{document}
\begin{frontmatter}


\title{Influence Diffusion in Social Networks under Time Window Constraints\tnoteref{label1}}
\tnotetext[label1]{An extended abstract of a preliminary version of this paper appeared
in: Proceedings of 20th International Colloquium on Structural Information and Communication Complexity (Sirocco 2013),
Lectures Notes in Computer Science vol. 8179, T. Moscibroda and A.A. Rescigno (Eds.), pp. 141-152, 2013. 
Parts of this work 
were done while the first 
and last author were visiting
Simon Fraser University and while the third author was visiting the
University of Salerno.
This research was supported in part by the Ebco Eppich Endowment Fund at
Simon Fraser University and by NSERC of Canada.}

\author[1]{Luisa Gargano}

\author[2]{Pavol Hell}

\author[2]{Joseph G. Peters}

\author[1]{Ugo Vaccaro}

\address[1]{Dipartimento di Informatica, University of Salerno, Italy}

\address[2]{School of Computing Science, Simon Fraser University,  Canada}

\begin{abstract}
We study a combinatorial model of the spread of influence   in networks
that generalizes existing schemata recently proposed in the literature.
In our model, agents  change behaviors/opinions on the basis
of information collected from their neighbors  in a time interval of \emph{bounded size} 
whereas agents are assumed to have \emph{unbounded memory} in previously studied  scenarios.
In our mathematical  framework, one is given a network $G=(V,E)$, an integer
value $t(v)$ for each node $v\in V$, and
a time window size $\lambda$.
The goal is to determine a small
set of nodes (\emph{target set}) that
influences the whole graph. The spread of influence proceeds in rounds
as follows: initially all
nodes in the target set are influenced; subsequently, in each
round, any uninfluenced node $v$ becomes influenced if the number
of its neighbors that have been influenced in the previous
$\lambda$ rounds is greater than or equal to $t(v)$.
We prove that the problem of finding a minimum cardinality
target  set that influences the whole network $G$ is hard to approximate
within a polylogarithmic factor. On the positive side, we design
exact polynomial time algorithms for paths, rings, trees, and complete graphs.
\end{abstract}

\begin{keyword}
Target set selection, Influence diffusion
\end{keyword}

\end{frontmatter}
%

%

\section{Introduction}
Many phenomena can be represented by dynamical processes on networks. 
Examples include cascading failures in physical infrastructure networks \cite{DOV},
information cascades in social and economic systems \cite{BHW},
spreads of infectious diseases \cite{AM}, and the
spreading of ideas, fashions, or behaviors among people \cite{CM,UBMK}.
Therefore, it comes as no surprise that the study of dynamical 
processes on complex networks is an active area of research, crossing a variety of 
different disciplines.
Epidemiologists, social scientists, physicists, and computer scientists have studied 
diffusion phenomena using very similar models to
describe the spreading of diseases, knowledge, behaviors, and innovations 
among individuals of a population (see \cite{BBV,BLMCH,EK} for surveys of the area).

A particularly important diffusion process 
is that of \emph{viral marketing} \cite{LAM}, which
refers to the spread of information about products and  behaviors  and their
adoption by people. Recently, it has also become  an important tool in
the communication strategies of politicians \cite{LKLG,T}
(see also \cite{RW} for a nice survey of the area).
Although there are many similarities between social 
and epidemiological contagion  \cite{DW},
social contagion is usually an intentional act
on the part of the
transmitter and/or the adopter, unlike a pathogen
contagion. The spread of ideas requires  extra  mechanisms in addition to mere
exposure, e.g.,   some kind of ``social pressure''. 
More importantly, in the  marketing scenario one is interested in
\emph{maximizing} the spread of information \cite{DR-01}, while this is not
likely to happen in the spread of pathogenic viruses.
The intent of maximizing the spread of viral information across a network naturally
suggests many optimization  problems. Some of them  were first articulated in the seminal papers
\cite{KKT-03,KKT-05}, under various adoption paradigms.
The recent monograph \cite{CLC} contains an excellent  
description of the area.
In the next section, we will explain and motivate our model of 
information diffusion, state the problem that we are investigating,
describe our results, and discuss how
they relate to the existing literature.

\section{The Model, the Context, and the  Results}

The network is represented by a pair $(G,t)$, where $G=(V,E)$ is an undirected  graph and 
$t: V \longrightarrow \N = \{1,2,\ldots, \}$ is a
function assigning integer thresholds to  nodes. 
We assume that $1\leq t(v)\leq \deg(v)$ for each  $v\in V$,
where $deg(v)$ is the degree of $v$.
For  a  given set $\SS\subseteq V$ and a time window size $\lambda\in \N$, 
we consider a  dynamical  process of influence diffusion in $G$ defined
by two sequences of node subsets,
$\influenced[\SS,r]$ and $\active[\SS,r]$, $r=0, 1, \ldots ,$ where

$$\influenced[\SS,0]=S, \qquad \active[\SS,0]=\emptyset,$$
and for any $r\geq 1$ it holds that
\begin{eqnarray}\label{eq-def}
\influenced[\SS,\i]&=&\influenced[\SS,\i-1]\cup \left\{v:\big|N(v)\cap \active[\SS,\i]\big|\geq t(v)\right\}\;\;\;\;\;\nonumber\\
&&\\
  \active[\SS,\i]&=&\begin{cases}{\influenced[\SS,\i-1]}& {\mbox{ if $r \leq \lambda$}}\\
                                     {\influenced[\SS,\i-1]\setminus \influenced[\SS,\i-1-\lambda]}& 
                                     {\mbox{ if $r> \lambda$}}\nonumber
												\end{cases}
\end{eqnarray}
Intuitively, the set $S$ might  represent a group of people who are initially 
influenced/convinced  to adopt a product or an idea.
Then the cascade proceeds in rounds. In each round $\i$, the set 
of influenced nodes is
augmented by including each node $v$ that has a number of influenced 
and \emph{still active}
neighbors   greater than or equal to its threshold $t(v)$. 
A node is active for $\lambda$ rounds after it
becomes influenced and then it becomes inactive.
 
Our model is based on the models in~\cite{DK,MT} which
assume that people can be divided into three classes at any time instant.
\emph{Ignorants} are those not aware of a rumor/not yet influenced, 
\emph{spreaders} are those who are spreading it, and \emph{stiflers} are
those who know the rumor/have been influenced but have ceased to
spread the rumor/influence.\footnote{The reader will notice an analogy 
with the well known SIR model of mathematical epidemiology \cite{AM}, in 
which individuals can be classified as  Susceptible, Infected, and Recovered.}
Several rules have been proposed to 
govern the transition from ignorants to spreaders and 
from spreaders to stiflers, 
and many papers have  studied the dynamics of these systems, 
mostly in stochastic scenarios (see \cite{BCKC,NMBM} and references quoted therein).
Here, we  posit  that any ignorant node  becomes a spreader if the number of its
neighbors  who are spreaders is
 above a certain threshold  (i.e., the node is subject to a large enough
amount of  ``social pressure''), and any spreader becomes a stifler after 
$\lambda$ rounds
(because the spreader loses interest in the rumor, for instance).
Other papers  have studied information diffusion under  
similar assumptions  \cite{CMP,KH}.

Our model also captures
another important characteristic of influence diffusion.
Indeed, research in Behavioural Economics 
shows that humans take decisions mostly on the basis of very recent events, 
even though they might hold much more in their memory \cite{Alba,Chen+}.
Moreover, It is well known (e.g. \cite{BBM}) that people are more
inclined  to react to pieces of information cumulatively heard  during a
``short'' time  interval  than to information heard during a considerably longer period of time.
In other words, one is more likely to be convinced of 
an opinion heard from  a certain number  of friends during  the last
few days than by an opinion heard sporadically during the last year from the \emph{same} number of people. 
Therefore, it seems reasonable
to study diffusion processes in which people  have
\emph{bounded memory}, and only the number of spreaders
heard during the last $\lambda$  rounds may contribute to the 
change of status of an ignorant node.\footnote{Another model in which
individuals carry a memory of the ``amount of influence'' received 
during a bounded time interval has been 
studied in \cite{DW}.}
Formally, 
one has    a  dynamical  process of influence diffusion   on $G$ 
described  
by the  sequence of node subsets
$\influenced'[\SS,r]$, $r=0, 1, \ldots ,$ where
$\influenced'[\SS,0]=S$,
and for any $r\geq 1$ it holds that
\begin{equation}\label{eq-def2}
\influenced'[\SS,\i]=\influenced[\SS,\i-1]\cup \left\{v:\big|N(v)\cap \influenced'[\SS,\i-1]\big|\geq t(v)\right\}
\end{equation}
if $r\leq \lambda$, and 
\begin{eqnarray}\label{eq-def3}
\influenced&\!\!\!'\!\!\!&[\SS,\i]=\influenced'[\SS,\i-1]\\
                    &\!\!\!\cup\!\!&
\left\{v:\!\big|N(v)\cap (\influenced'[\SS,\i-1]\!\setminus\! \influenced'[\SS,\i-1-\lambda])\big|\geq t(v)\right\}\nonumber
\end{eqnarray}
if $r> \lambda$.

It is immediate that 
(\ref{eq-def2}) and (\ref{eq-def3}) are an equivalent  way
to write (1) and (2):
  for any $S\subseteq V$ and $r\geq 1$,
$\influenced'[\SS,\i]=\influenced[\SS,\i],$
so we get that  the spreading process with ``stiflers'' also describes
the spreading process with ``bounded memory'' governed   by (\ref{eq-def2}) and (\ref{eq-def3}).

 \smallskip
\noindent
Summarizing, 
the  specific problem that
we  shall  study in this paper 
 is the following:

\smallskip
\noindent
\textsc{Time Window Constrained Target Set Selection} (TWC--TSS)\\
{\bf Input:} A graph $G=(V,E)$, a threshold function 
$t:V\longrightarrow \mathbb{N}$, and a \span size $\lambda$.\\
{\bf Output:} A minimum size $\SS\subseteq V$ such that 
$\influenced[\SS,\i]=V$, for some $\i\geq 0$.

\smallskip
When $\lambda$ is large enough, for instance equal to
the number $n$ of nodes, our \textsc{Time Window Constrained 
Target Set Selection}
problem is equivalent to the classical \textsc{Target Set Selection} 
problem 
studied in \cite{ABW-10,BCNS,BHLM-11,C+,Chen-09,Chiang,Chopin-12,Chun,Chun2,Cic+,C-OFKR,Re,Za},
among the others. 
Strictly related is also 
the area of  dynamic monopolies (see \cite{FKRRS-2003,Peleg-02}, for instance).
In terms of our second formulation of the TWC--TSS problem, 
the classical \textsc{Target Set Selection} problem can be viewed as an
extreme case in which it is assumed that people have unbounded memory.
In  general, the TWC--TSS and the TSS problems are quite different. One of the  
main  difficulties
of the new  TWC--TSS problem is that the sequence of sets $\active[\SS,\i]$,
$r=0,1,\ldots$ is not necessarily monotonically non-decreasing: it is possible that 
$\active[\SS,\i]$ is larger than $\active[\SS,\i+1]$ for some values of $r$.
When $\lambda=n$, we  have $\active[\SS,\i]=\influenced[\SS,\i-1]$ for any $r$,
and  monotonicity is restored. 
At the other extreme, when $\lambda=1$,
a node $v$ becomes  influenced at time $r$ only if at least $t(v)$ of its 
neighbors 
become influenced at \emph{exactly} time $r-1$.
This sort of synchronization in the propagation  of
influence   poses new 
challenges, both in the assessment of the computational 
complexity of the TWC--TSS problem and, especially,  in the design
of algorithms for its solution.
The example in which the graph $G$ is a path
is  particularly illuminating. As we shall see in Section \ref{sec:path},
the \textsc{Target Set Selection} problem is trivial  to solve on a path; 
it is far  from being so  when there is a fixed time window size $\lambda$.

\begin{example}\label{ex-1a}
Consider the tree $T$  in Figure \ref{fig1}.
The number inside each circle is the  node threshold.
For convenience, we consider $T$ to be rooted at $v_0$.
If the time window size is $\lambda =1$, then
one  set that  influences all nodes of  $T$   is 
$\SS=\{ v_{1,2},\, v_{2,2},\, v_{2,6}, \,v_{3,4}\}$.
Indeed we have
{\small{
\begin{eqnarray*}
 \influenced[\SS,0] &=& \{ v_{1,2}, v_{2,2}, v_{2,6}, v_{3,4}\}=\active[\SS,1]
 \\
 \influenced[\SS,1]&=&\influenced[\SS,0]\cup\{v_{1,1},v_{1,3},v_{2,1},v_{2,3},v_{2,5},v_{3,3},v_{3,5}\}
 \\
 \active[\SS,2]&=&\influenced[\SS,1]\setminus\influenced[\SS,0]=\{v_{1,1},v_{1,3},v_{2,1},v_{2,3},v_{2,5},v_{3,3},v_{3,5},\}\\
 \influenced[\SS,2]&=&\influenced[\SS,1]\cup \{  v_0, v_{2,4}, v_{3,2}  \}   
 \\
 \active[\SS,3]&=&\influenced[\SS,2]\setminus \influenced[\SS,1]=\{ v_0, v_{2,4}, v_{3,2} \} 
 \\
 \influenced[\SS,3]&=&\influenced[\SS,2]\cup\{v_{3,1}\}=V
\end{eqnarray*}
}}
In particular we notice  that 

-- $v_{2,4}\in\influenced[\SS,2]$ since both 
its child  $v_{2,5}$ and its parent  $v_{2,3}$ are active  in round 2;

--  $v_0\in \influenced[\SS,2]$ since $t(v_0)=2$ of 
its children are active in  round 2;
 
-- $v_{3,1}\in\influenced[\SS,3]$ since  both the root 
$v_0$ and its child $v_{3,2}$ are active  in round 3;
 

\begin{figure}[ht!]\label{fig1}
\includegraphics[height=8.5truecm,width=10.7truecm]{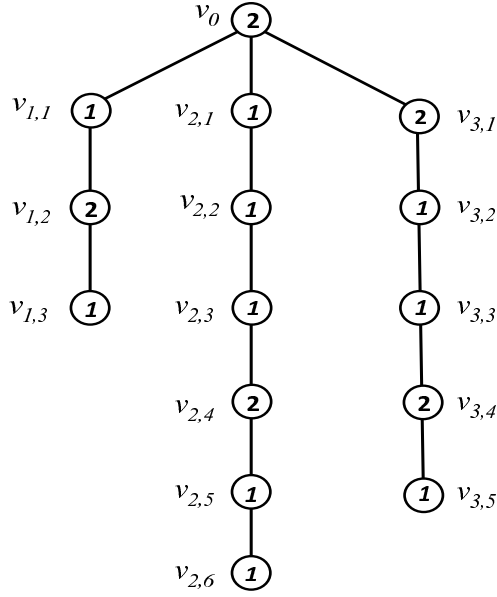}
\caption{The tree $T$, rooted at node $v_0$, considered in Example \ref{ex-1a}.
The number inside each node is its  threshold.}
\end{figure}
%
\noindent
If we assume a larger value of $\lambda$, then a smaller set can be choosen;  in particular if   $\lambda\geq 3$ then 
the set 
$\SS=\{v_{1,2}, v_{2,4},  v_{3,4}\}$  influences all nodes of  $T$. Indeed for $\lambda= 3$ we have
{\small{
\begin{eqnarray*}
 \influenced[\SS,0] &=& \{ v_{1,2}, v_{2,4},  v_{3,4}\}=\active[\SS,1]
 \\
 \influenced[\SS,1]&=&\influenced[\SS,0]\cup\{v_{1,1},v_{1,3},v_{2,3},v_{2,5},v_{3,3},v_{3,5}\}=\active[\SS,2]
 \\
 \influenced[\SS,2]&=&\influenced[\SS,1]\cup \{  v_{2,2},v_{2,6},  v_{3,2}  \} \\
 \active[\SS,3]&=&\influenced[\SS,2]= \{v_{1,1},v_{1,2},v_{1,3},v_{2,2},v_{2,3},v_{2,4}, v_{2,5},v_{2,6},v_{3,2},v_{3,3},v_{3,4},v_{3,5}\} 
 \\
 \influenced[\SS,3]&=&\influenced[\SS,2]\cup \{  v_{2,1} \} \\
\active[\SS,4]&=&\influenced[\SS,3] \setminus \influenced[\SS,0]= 
                              \{v_{1,1},v_{1,3},v_{2,1},v_{2,2},v_{2,3}, v_{2,5},v_{2,6},v_{3,2},v_{3,3},v_{3,5}\} 
																\\
 \influenced[\SS,4]&=&\influenced[\SS,3]\cup \{  v_{0} \} 
                                = \{v_{0},v_{1,1},v_{1,2},v_{1,3},v_{2,1},v_{2,2},v_{2,3},v_{2,4}, v_{2,5},v_{2,6},v_{3,2},v_{3,3},v_{3,4},v_{3,5}\} 
																\\
\active[\SS,5]&=&\influenced[\SS,4]	\setminus \influenced[\SS,1]=\{v_{0},v_{2,1},v_{2,2},v_{2,6},v_{3,2}\} 
																\\
\influenced[\SS,5]&=&\influenced[\SS,4]\cup \{  v_{3,1} \} 
                                = V												
\end{eqnarray*}
}}
\end{example}

\medskip
\noindent
\textbf{Our Results.}
In Section \ref{sec:hard}, we prove a polylogarithmic inapproximability result 
for the TWC--TSS problem under a plausible computational complexity assumption. The result is obtained
by a modification of 
a proof of the inapproximability of TSS by Chen
\cite{Chen-09}. 
In view of the strong  inapproximability 
 of the TWC--TSS problem, 
  we then turn our attention  to special cases of the problem.
  In Section \ref{sec:poly} we present the main results of the paper:
 exact polynomial time algorithms for paths, rings, complete graphs, and trees.
 The algorithms for paths and rings are based on dynamic programming, and 
 the algorithm for complete graphs is greedy.
 The algorithm for trees is also  based on  dynamic programming  and  
 requires the solution of 
 polynomially many integer linear programs. 
 The polynomial time solvability of each
 integer linear program is
 guaranteed by the
 unimodularity of
 the associated matrix of coefficients.

\section{Hardness of \PROB}\label{sec:hard}

In general, our optimization problem \PROB is unlikely to be
efficiently  approximable, as the following result shows.

The following theorem is a generalization  of a similar inapproximability result
  given in \cite{Chen-09} for the 
  \textsc{Target Set Selection} problem that, as said before, 
 corresponds to our \textsc{Time Window Constrained Target Set Selection}
 problem when the time window size  $\lambda$ is unbounded.
 Our result holds for any fixed value of $\lambda$.

\begin{theorem}\label{teo1}
For any fixed value of the time window size $\lambda$,
the \PROB problem cannot be approximated within a ratio 
of  $O(2^{\log^{1-\epsilon} n})$ for any fixed $\epsilon>0$, 
unless $NP\subseteq DTIME(n^{polylog(n)})$.
\end{theorem}

\proof 
We prove the theorem   by   a polynomial time reduction from 
the same \minrep  problem used in \cite{Chen-09}.
Let $G=(V,E)$ be a bipartite graph, where $V=V_A\cup V_B$,
$V_A\cap V_B=\emptyset$, and  $E\subseteq V_A\times V_B$. 
Let    $\cA$ be a family of subsets of $V_A$ that  partitions  
$V_A$  into $|\cA|$ equally sized subsets,  
and analogously let the family $\cB$ be 
 a partition of $V_B$  into $|\cB|$ equally sized subsets.
Given  graph $H$ and   partitions $\cA$, $\cB$, the \minrep
problem asks for a subset  $U\subseteq V$ of minimum size  such that 
for each $A\in\cA$ and $B\in\cB$ 
\begin{equation}\label{eq-cap}
 E\cap (A\times B)\neq \emptyset \mbox{ implies } 
 [E\cap  (A\times  B)]\cap  (U\times  U)\neq \emptyset.
\end{equation}

Given an instance ${\cal I}$ of \minrep  consisting of the bipartite graph $G=(V,E)$
and  the pair of  partitions ($\cA,\cB$),
  we construct an instance ${\cal I'}$ of the \PROB problem.
  More precisely, 
we will construct a suitable graph $G'=(V',E')$ and threshold function $t: V' \longrightarrow \N = \{1,2,\ldots, \}$,
but we will not fix $\lambda$  since our aim is to prove inapproximability for \emph{any} 
  value of $\lambda$.



\bigskip

\begin{figure}
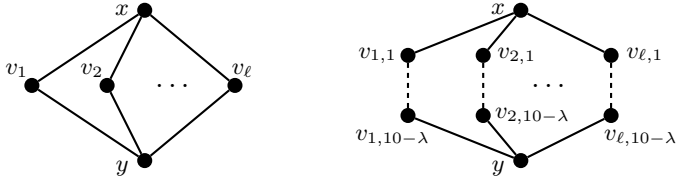
\label{figure2}

\cnode[fillstyle=solid, fillcolor=black](3,0){.1cm}{x}
\rput(2.7,0){\mbox{\footnotesize $x$}}
\cnode[fillstyle=solid, fillcolor=black](1.5,-1){.1cm}{1}
\cnode[fillstyle=solid, fillcolor=black](2.5,-1){.1cm}{2}
\cnode[fillstyle=solid, fillcolor=black](4.2,-1){.1cm}{3}
\rput(3.4,-1){\ldots}
\cnode[fillstyle=solid, fillcolor=black](3,-2){.1cm}{y}
\rput(2.7,-2.1){\mbox{\footnotesize $y$}}
\ncline{x}{1}
\ncline{x}{2}
\ncline{x}{3}
\ncline{3}{y}
\ncline{2}{y}
\ncline{1}{y}
\rput(1.3,-.8){\mbox{\footnotesize $v_1$}}
\rput(2.3,-.8){\mbox{\footnotesize $v_2$}}
\rput(4.3,-.8){\mbox{\footnotesize $v_\ell$}}
\cnode[fillstyle=solid, fillcolor=black](8,0){.1cm}{xb}
\rput(7.7,.02){\mbox{\footnotesize $x$}}
\cnode[fillstyle=solid, fillcolor=black](6.5,-.6){.1cm}{1b}
\cnode[fillstyle=solid, fillcolor=black](7.5,-.6){.1cm}{2b}
\cnode[fillstyle=solid, fillcolor=black](9.2,-.6){.1cm}{3b}
\cnode[fillstyle=solid, fillcolor=black](6.5,-1.4){.1cm}{1c}
\cnode[fillstyle=solid, fillcolor=black](7.5,-1.4){.1cm}{2c}
\cnode[fillstyle=solid, fillcolor=black](9.2,-1.4){.1cm}{3c}
\rput(8.4,-1){\ldots}
\cnode[fillstyle=solid, fillcolor=black](8,-2){.1cm}{yb}
\rput(7.7,-2.1){\mbox{\footnotesize $y$}}
\ncline{xb}{1b}
\ncline{xb}{2b}
\ncline{xb}{3b}
\ncline[linestyle=dashed, dash=2pt 2pt]{1c}{1b}
\ncline[linestyle=dashed, dash=2pt 2pt]{2c}{2b}
\ncline[linestyle=dashed, dash=2pt 2pt]{3c}{3b}
\ncline{3c}{yb}
\ncline{2c}{yb}
\ncline{1c}{yb}
\rput(6.1,-.6){\mbox{\footnotesize $v_{1,1}$}}
\rput(7.95,-.61){\mbox{\footnotesize $v_{2,1}$}}
\rput(9.65,-.6){\mbox{\footnotesize $v_{\ell,1}$}}
\rput(6.3,-1.7){\mbox{\footnotesize $v_{1,10-\lambda}$}}
\rput(8.16,-1.4){\mbox{\footnotesize $v_{2,10-\lambda}$}}
\rput(9.6,-1.7){\mbox{\footnotesize $v_{\ell,10-\lambda}$}}

\bigskip

\bigskip

\bigskip

\bigskip

\bigskip

\caption{ (a) The gadget $\Gamma_\ell$ consisting of 
$\ell$ paths of length 2 sharing the extremal nodes.
 (b) The gadget $\Gamma^\lambda_\ell$ 
consisting of  $\ell$ paths of length $ 11-\lambda$ sharing the extremal nodes.}
\end{figure}

We denote by $\Gamma_\ell$ the gadget shown in Figure \ref{figure2}(a), which consists of $\ell$ paths of length 2 
connecting the same pair of nodes. If $\lambda \leq 8$, we need another gadget   $\Gamma^\lambda_\ell$  
which is shown in Figure \ref{figure2} (b); it  consists of $\ell$ paths, each having  length   $ 11-\lambda$  and
connecting the same pair of extremal nodes.
All internal nodes of the gadgets have threshold 1.

\medskip

Now, given an instance $\cal I$ of {\tt MIN$\,$REP}, we construct an instance
$\cal I'$ of \PROB as follows.
Let $N=|V|+|E|$. 
The graph $G'$
consists of the node sets $V_1, V_2, V_3,V_4$  and the connecting gadgets  as follows:
\begin{itemize}
\item  $V_1=V$ and each node has threshold $N^2$,
\item $V_2=\{x_{(a,b)}  :  (a,b)\in E\}$;   each node $x_{(a,b)}\in V_2$  has threshold $2N^5$ and is connected to each of  $a\in V_1$ and $b\in V_1$ by 
 a gadget $\Gamma_{N^5}$.
If $\lambda \leq 8$, each $x_{(a,b)}$ is also connected to each of $a\in V_1$ and $b\in V_1$  by a gadget $\Gamma^\lambda_{N^5}$.
\item $V_3=\{y_{A,B} : A\in\cA, B\in\cB, (A\times B)\cap E\neq\emptyset\}$;  each node $y_{A,B}\in V_3$  has threshold $N^4$ and is connected   by a gadget $\Gamma_{N^4}$ to each  $x_{(a,b)}\in V_2 $ with $a\in A$ and $b\in B$, and
\item $V_4=\{z_1,\ldots, z_N\}$; each node $z\in V_4$  has threshold $|V_3| N^2$ and is connected   by a gadget $\Gamma_{N^2}$ to each  node in $ V_3 $  and by a gadget $\Gamma_N$ to each node in $V_1$.
\end{itemize}

We shall show that the size of any optimal solution for the \minrep instance
$\cal I$ is at most twice the size of
an optimal solution for the \PROB instance $\cal I'$.
\\ 
First, suppose that $U$ is  an optimal \minrep solution for $\cal I$ and consider  $U\subseteq V_1$ in the \PROB instance $\cal I'$.
Since $U$ is a \minrep solution, we know that for each $A\in\cA$ and $B\in\cB$ it holds that
\begin{equation}\label{eq-E}
 E\cap (A\times B)\neq \emptyset \quad \Rightarrow \quad E\cap  ((A\cap U)\times  (B\cap U))\neq \emptyset.
 \end{equation}
Now recall the existence of the gadgets $\Gamma_{N^5}$   between nodes in $V_1$ and nodes in $V_2$ and that each node in $V_2$
has threshold $2N^5$. From this and (\ref{eq-E}) we get that
 $\influenced[U,2]$ contains
 a subset of nodes in $V_2$ that can influence (through the $\Gamma_{N^4}$  gadgets between nodes in $V_2$ and $V_3$) 
 each node in $V_3$ in round $4$, that is, 
 $$V_3\subseteq \influenced[U,4].$$
This allows all of the nodes in $V_4$ to become influenced in round 
$6$, that is,
$$V_4\subseteq \influenced[U,6].$$
 Hence,   $V_4\subseteq \active[U,7]$ and  the remaining nodes in $V_1$, namely $V_1-U$, become influenced in round $8$. 

Therefore,  all nodes in  $V_1$  have become influenced  
 either in round  0 or in round $8$ and
$$V_1\subseteq 
     (\active[U,1]\cup\ldots\cup\active[U,\lambda])\cup\active[U,9].$$
The influence of the nodes in $V_1-U$ reaches the nodes in $V_2$ in round 10
through $\Gamma_{N^5}$ gadgets. If $\lambda >8$, then the nodes in
$U\subseteq V_1$ are still active ib round 9 and can influence nodes in $V_2$
in round 10 through $\Gamma_{N^5}$ gadgets.  If $\lambda \leq 8$, then the
influence of the nodes in $U$ reaches the nodes in $V_2$ in round $11-\lambda$
through $\Gamma^\lambda_{N^5}$ gadgets and this influence remains until round
$11-\lambda +(\lambda -1) = 10$. In either case,
each uninfluenced node $x_{(a,b)}\in V_2$ has at least $N^5$  neighbors 
 that belong to $\active[U,10]$ for both $a,b\in V_1$.
Hence,  $V_2\subseteq \influenced[U,10]$  thus
completing the  process.

\medskip

We want to show now that if we have a solution $\SS$ to TWC--TSS, then we can construct a 
solution to \minrep of size at most $2|\SS|$.

{We want to see now that we can  assume $\SS\subset V_1\cup V_4$. To this aim, we first  notice that the influence process can only proceed from nodes in $V_1$ to nodes in $V_2$, from nodes in $V_2$  to nodes in $V_3$, and  from nodes in $V_3$  to nodes in $V_4$; namely }we have that: 
\begin{itemize}
\item[a)] 
A node in $V_4$ can become influenced at  some round $\i$ only if {\em all}  nodes 
in $V_3$ are active in round $\i-{2}$ (otherwise the threshold $|V_3|N^2$ cannot be reached,
irrespectively of which nodes in $V_1$ are active; indeed if there exists a node in $V_3$ which is not active at round $r-2$, then the sizes of the gadgets between a node   $z\in V_4$ and  nodes in  $V_3$ and $V_1$ implies that at round $\i-1$ at most $(|V_3|-1)N^2+|V_1|N< |V_3| N^2$ neighbors of $z$ can be active).
\item[b)]  
A node  $y_{A,B}\in V_3$ can become influenced at  some round $\i$ only if  there exist $a\in A$ and $b\in B$ such that  the node $x_{(a,b)}\in V_2$ is active in round $\i-{2}$ (otherwise the threshold $N^4$ cannot be reached
irrespectively of which nodes in $V_4$ are active).
\item[c)]  
A node in $x_{(a,b)}\in V_2$ can become influenced in some round $\i$ only if  both $a$ and $b$
($a,b\in V_1$) are  active in round $\i-{2}$ 
{if $\lambda >8$ and in round $\i-2$ or $\i-(11-\lambda)$ if $\lambda\leq 8$ }(otherwise the threshold $2N^5$ cannot be reached
irrespectively of which nodes in $V_3$ are active).
\end{itemize}

\noindent 
{Moreover,  we can assume that:}
\begin{itemize}
\item
$\SS$ does not contain any node internal to 
any gadget; indeed, due to the  thresholds of the extremal nodes of the gadgets, such nodes  could be useful only if $\ell$   nodes in a gadget  $\Gamma_\ell$  (or $\Gamma^\lambda_\ell$)
  are in $\SS$, but any gadget has $\ell\geq N>|V|$.
\item
$\SS\cap V_3=\emptyset$; indeed, {by b) and a)} 
we can replace each  $y_{A,B}\in \SS\cap V_3$
by  a node $x_{(a,b)}\in V_2\cap E$ for some $a\in A$ and $b\in B$.
\item
 $\SS\cap V_2=\emptyset$; indeed, recalling that  we want a solution that is upper bounded by $2|\SS|$, {using c)} we can replace each  $x_{(a,b)}\in \SS\cap V_2$ by  the two nodes, $a$ and $b$ in $V_1$.
\end{itemize}
Summarizing, we can  assume $\SS\subset V_1\cup V_4$. However,  
nodes in $\SS\cap V_4$ cannot influence any node in $V_1-\SS$.
By a), b), and c), this implies that $\SS\cap V_1$ can influence 
all nodes in $V'$ and is a solution for the \minrep instance $\cal I$.
\hfill $\Box$

\section{Polynomially Solvable  Cases of \PROB}\label{sec:poly}

We now   present exact polynomial time algorithms 
to  solve the \PROB problem in several classes of graphs.

\subsection{Paths}\label{sec:path}

Let $L^n=(V,E)$ be a path on $n$ nodes, with $V=\{0,\ldots, n-1\}$ 
and $E=\{(i,i+1)\ :\ 0\leq i\leq n-2\}$.
 Since the threshold of each node cannot exceed its degree, 
 we have that $t(0)=t(n-1)=1$ and $t(i)\in\{1,2\}$, for each $i=1,\ldots, n-2$.  
 
 The \PROB problem is trivial to solve in case $\lambda$ is unbounded. 
 Letting
 $\{{i_1}, {i_2}, \ldots , {i_m}\}$ be the nodes of $L^n$ having 
 threshold equal to $2$, 
 one can see that 
 $\{{i_1}, {i_3}, \ldots ,{i_{m-2}},  {i_{m}}\}$ is 
 an optimal solution 
when  $m$ is odd,
 whereas the subset   $\{{i_1}, {i_3}, \ldots , {i_{m-1}},  {i_{m}}\}$ is optimal  when $m$ is even.
 In case $\lambda$ has some fixed value, 
 the situation is much  more complicated. 
Indeed, because of the time window constraint,
 one must judiciously choose the initial target set  in such a way
 that, for every   node with threshold 2 that does not belong to the initial target set, 
its two neighbors become
influenced at the correct times.

To avoid trivialities,   we 
assume that  $L^n$ has  at least two nodes with threshold equal to 2.
 Should it be otherwise, for instance 
  all  nodes have threshold 1,  then any   subset $S$ of $V$ with $|S|=1$ 
 is an optimal solution.
If exactly one node, say  $i$, has threshold 2, then  $\{i\}$ is  
an  optimal solution.

\begin{lemma}\label{claim-1}
If $\ell=\min\{i\in V :  t(i)=2\}$ and $s=\max\{i\in V :  t(i)=2\}$, 
then there exists an optimal solution  $\SS$ such that
\begin{itemize}
\item[a)] $\SS\cap \{0,\ldots \ell-1\}=\emptyset= \SS\cap \{s+1,\ldots n-1\}$;
\item[b)]  $\ell, s \in \SS$.
\end{itemize}
\end{lemma}

\proof  
Since $t(\ell)=2=t(s)$, we immediately get that 
\begin{equation}\label{eq-ls}
\SS\cap \{0,\ldots ,\ell\}\neq \emptyset \mbox{ and } \SS\cap \{s,\ldots ,n-1\}\neq \emptyset.
\end{equation}
It is easy to see  that  if $i\in \SS$ with $i<\ell$ (resp.  $i>s$), 
then the set $\SS'=\SS-\{i\}\cup \{\ell\}$ 
(resp. $\SS'=\SS-\{i\}\cup \{s\}$) is also an optimal solution. Therefore,  
we can assume that $\SS$ satisfies a).
If $\SS$ is an optimal solution satisfying a), then a) and (\ref{eq-ls}) imply
that $\SS$ satisfies b). 
\hfill $\Box$

Lemma \ref{claim-1} implies that we can ignore 
   all nodes in $L^n$ that are either to the left of
  the lowest numbered node with threshold 2, or to the right of
  the highest numbered node with threshold 2. Equivalently, from  now on 
  we can assume that 
  $$t(0)=t(n-1)=2.$$
	
\noindent
For each $i=0,\ldots, n-1$, let   $L_i^n$ denote the sub-path 
consisting of  the last $n-i$ nodes
$\{i,i+1,\ldots, n-1\}$ of $L^n$. 
We denote by 
$\SN(i)$ the minimum size of a TWC target set
  for $L_i^n$  that contains both the extreme nodes, 
  that is,  $i$ and  $n-1$. 
  Our   goal is to compute $\SN(0)$, the size of an 
  optimal solution for $L_0^n=L^n$.

Let us 
   define the array 
$D[0\ldots (n-1)]$, where $D[n-1]=n-1$ and for each $0 \leq i <n-1$,
  \begin{equation}\label{eq-D}
  D[i]=\min\{j\ :\ i< j\leq n-1 \mbox{ and } t(j)=2\}.
  \end{equation}
 Since  $t(n-1)=2$, value $D[i]$ is always well defined.

One can  check that the following algorithm 
computes  an  array $D$ satisfying (\ref{eq-D}).

\medskip
\begin{center}
\begin{tabular}{l}
\hline
{\bf Algorithm ARRAY($L^n$)}  {\em [ Input: A path $L^n$ with threshold 
function $t(\cdot)$]}\\
{\bf $\ $1.} $j= n-1$\\
{\bf $\ $2.} $D[n-1]= n-1$\\
{\bf $\ $3.}
{\bf for} $i=(n-2)$ {\bf down to} 0 {\bf do} \\
{\bf $\ $4.}
\hphantom{F }    $D[i]= j$\\
{\bf $\ $5.}
\hphantom{F }   {\bf if}    $t(i)=2$  {\bf then}  $j= i$\\
\hline
\end{tabular}
\end{center}
\medskip
 \begin{lemma}\label{lemma-2i}
 Fix the time window size $\lambda$ and consider the family of all
  TWC target sets for $L_i^n$ that  
 include both node $i$ and node $n-1$. 
If $i <n-1$, then such a family contains a minimum size TWC target set
whose  second smallest element belongs to the set
\begin{equation}\label{eq-j}
\big\{D[i]\big\}\cup \big\{x  : \max\{D[i]+1, \, 2D[i]-i- \lambda +1\}\leq x \leq \min \{2D[i]-i+\lambda-1, \, D[D[i]]\}\big\}.
\end{equation}
\end{lemma}

\proof
Let $\SS$ denote  a minimum size solution for $L_i^n$ among all the TWC 
target sets containing both $i$ and $n-1$.
Clearly,  the smallest element in $\SS$ is $i$.
Let $j$ be the second smallest element of $\SS$ and 
assume by contradiction that $j$ does not belong to the set in (\ref{eq-j}). 
We distinguish four cases based on the value of $j$.

\medskip
\noindent
{{\bf Case}  $0<j < D[i]$:}
In this case, the set $\SS'=\SS-\{j\}\cup\{D[i]\}$ is  a TWC target 
set of size at most $|\SS|$ satisfying the claim. 
Indeed, by definition of the vector $D$, 
any node $x$, $i< x <D[i]$, has  $t(x)=1$. Since $i\in \SS$, we 
have that $x$ is influenced no later than round $x-i$, 
that is, $x\in \influenced[\SS',x-i]$. Moreover, the influencing of 
any node $y>D[i]$ 
cannot  depend on the presence of  $j$ in the target set, that is, 
$\SS-\{j\}=\SS'-\{D[i]\}$ is a TWC target set for 
$L_{D[i]+1}^{n-1}$.

 \medskip
 \noindent
{{\bf Case} $D[i]<j< 2D[i]-i-\lambda+1$:}
In this case, we have that $D[i]\notin \SS$ (since $0<D[i]<j$) 
and $t(D[i])=2$.  Therefore, to become
influenced, node $D[i]$ needs two neighbors that are active in the 
same round.
 Each node $x<D[i]$ becomes influenced thanks to $i\in \SS$, 
 namely $x\in\influenced[\SS,x-i]$. 
In particular, node $D[i]-1$ is influenced 
in round $D[i]-1-i$ and remains active for $\lambda$ rounds, 
that is,
\begin{eqnarray}
(D[i]-1)&\in&\bigcap_{\i=D[i]-i}^{D[i]-i+\lambda-1}\active[\SS,\i], \mbox{ and } \label{eq-D2a} \\
(D[i]-1)&\not\in& \active[\SS,r] \mbox{ for } r \leq D[i]-i-1 \mbox{ and } r\geq D[i]-i+\lambda. \label{eq-D2b}
\end{eqnarray}
Analogously,  each node $y$ with $D[i]<y<j$ becomes influenced 
thanks to $j\in \SS$, namely $y\in\influenced[\SS,j-y]$. 
In particular,
 node $D[i]+1$ is influenced in round $j- D[i]-1$ and remains 
 active for  $\lambda$  rounds, that is,    
\begin{equation}\label{eq-D4} 
(D[i]+1)\in\bigcap_{\i=j- D[i]}^{j- D[i]+\lambda-1}\active[\SS,\i] 
 \mbox{ and } (D[i]+1)\not\in \active[\SS,j- D[i]+\lambda].
\end{equation}
The  hypothesis  of this case implies that
$j- D[i]+\lambda-1< (2D[i]-i-\lambda+1) - D[i]+\lambda-1=D[i]-i.$
This inequality, together with  (\ref{eq-D2a}) and 
(\ref{eq-D2b}), imply that  there is no round $\i$ 
such that 
both $D[i]-1$ and $D[i]+1$ belong to $\active[\SS,\i]$; thus  
 node $D[i]$ cannot be influenced and we have reached a contradiction.
 
 \medskip
 \noindent
{{\bf Case} $j> 2D[i]-i+\lambda-1$:}
As already seen in the previous case,  node $D[i]-1$ becomes influenced 
in round $D[i]-1-i$ and remains active  for the subsequent $\lambda$  
rounds as stated in (\ref{eq-D2a}) and (\ref{eq-D2b}).
\\
On the other side, node $D[i]+1$ becomes influenced in round $j- D[i]-1$,  
and
$$
(D[i]+1)\in\active[\SS,j-D[i]]  \mbox{ but } (D[i]+1)\not\in\active[\SS,\i], \forall \i \leq j-D[i]-1.								
$$
The hypothesis of this case implies that $j-D[i]>(2D[i]-i+\lambda-1)-D[i]=D[i]-i+\lambda-1$. This, together with (\ref{eq-D2a}) and (\ref{eq-D2b}), imply
 that there is no round when both
neighbors of $D[i]$ are active; thus  
 node $D[i]$ cannot be influenced. Again a contradiction.

\medskip
 \noindent
{{\bf Case} $j> D[D[i]]$}:
We have that no node $x$ with $i<x<j$ belongs to $\SS$. In particular,
 $D[i]$ and $D[D[i]]$ are two nodes with threshold  2 such that   
 $i<D[i]<D[D[i]]<j$.
 This  clearly prevents the two nodes $D[i]$ and $D[D[i]]$ from 
 becoming influenced. \hfill $\Box$

\medskip
From Lemma \ref{lemma-2i}, we have  $\SN(n-1) = 1$  
and, for each $i=0,\ldots, n-2$,
  \begin{equation}\label{eq-Si}
  \SN(i) =   1+ \min\big\{\SN(D[i]), \ \min_j \SN(j)\big\}
   \end{equation}
   where $j$ satisfies 
   $\max\big\{D[i]+1, \, 2D[i]-i-\lambda+1  \big\}\leq j \leq \min\big \{2D[i]-i+\lambda-1, \, D[D[i]]\big\}$.

\remove{
\begin{lemma}\label{lemma-line}
Let $\k_1=0<\k_2<\ldots < \k_m=(n-1)$ be   the nodes of $L^n$ with threshold 2.
For each $\ell=1,\ldots, m$ there exists 
an integer $\SN_\ell$ such that  for  each 
$i=\k_\ell, \ldots,\k_{\ell+1}$ the function $\SN(i)$
can only get one of the two values $\SN_\ell$ and $\SN_\ell -1$.
   \end{lemma}
	}
	\begin{lemma}\label{lemma-line}
For each node $\k\in\{0,\ldots,n-1\}$ of $L^n$  with $t(\k)=2$ there exists 
an integer $\SN_\k$ such that  for  each 
$i=\k, \ldots,D[\k]$ it holds $\SN(i)\in\{\SN_\k,\, \SN_\k -1\}$.
   \end{lemma}
	\proof  
	Let $\k_1=0<\k_2<\ldots < \k_m=(n-1)$ be   the nodes of $L^n$ with threshold 2.
	For each $\ell=2,\ldots, m$
	and  $i=\k_{\ell-1}, \ldots,  \k_{\ell}-1$,   we have that $D[i]=D[\k_{\ell-1}]=\k_\ell$ and
the value of the function $\SN(i)$ is obtained as in (\ref{eq-Si}) where the minimum is computed in the range
	\begin{equation}\label{eq-Si1}
	\max\big\{\k_\ell+1, \, 2\k_\ell-i-\lambda+1  \big\}\leq j \leq \min\big \{2\k_\ell-i+\lambda-1, \, \k_{\ell+1}\big\}.
	\end{equation}
	By definition we have  $\SN(\k_m)=\SN(n-1)= 1$ and we can fix $\SN_{\k_m}=1$.
	We prove now, by induction
on $\ell$ from $m-1$ down to $1$, that there exists a value $\SN_{\k_\ell}$ such that 
\begin{equation}\label{SN}
  \SN_{\k_\ell} -1\leq \SN(i)\leq \SN_{\k_\ell}\ 
\end{equation}
 for each integer $i=\k_\ell, \ldots,  \k_{\ell+1}$.
\\
By (\ref{eq-Si}), we have that $\SN(i)=\SN(\k_m)+1=2$
 for each $i=\k_{m-1},\ldots, \k_m-1$.
	Hence we can fix 
	$s_{\k_{m-1}}=2$
	and (\ref{SN}) holds for  $\ell=m-1$.
	
	\noindent
	Now suppose that (\ref{SN}) is true for each integer from $\ell$ to $m$; we prove it true for 
	$\ell-1\leq m-2$.
\\
	By  (\ref{eq-Si1}) and using  the inductive hypothesis on $\ell$, we have
	 $$\SN_{\k_\ell} \leq \SN(i)\leq  \SN_{\k_\ell}+1,$$
	for any  $i=\k_{\ell-1}, \ldots,  \k_{\ell}-1$.
	It follows that  the value 
	$\SN_{\k_{\ell-1}}=\max\{\SN(i)\ |\  \k_{\ell-1}\leq i\leq  \k_\ell-1\}$
	 satisfies (\ref{SN})  for $i=\k_{\ell-1},\ldots,  \k_\ell-1$.
	\\
	It remains to show that $\SN_{\k_{\ell-1}}-1\leq \SN(\k_\ell)\leq \SN_{\k_{\ell-1}}$.
	We notice that 
	 if $\SN(\k_\ell)=\SN_{\k_\ell}-1$, then by (\ref{eq-Si1}) we have  $\SN(i)=\SN(\k_\ell)+1=\SN_{\k_\ell}$ for each 
	$i=\k_{\ell-1}, \ldots,  \k_{\ell}-1$. Hence, we have that  $\SN_{\k_{\ell-1}}=\SN_{\k_\ell}=\SN(\k_\ell)+1$ and 
	(\ref{SN}) holds  for $i=\k_{\ell}$.  \hfill$\Box$

   \begin{theorem}\label{teo-line}
   For any \span size
 $\lambda$, 
 an optimal TWC target set for the path $L^n$  can be computed in time $O(n)$. 
   \end{theorem}

\proof The  size of an optimal target set for $L_n$ can  be computed  as $\SN(0)$. 
%
By Lemma \ref{lemma-line},
  one can implement  the computation in (\ref{eq-Si1}) as shown in the following  algorithm LINE($L^n$). 
	\\
	In the algorithm,  the variable $\k$   represents the  current node of threshold 2 
	and  the variable $\SN$ has value $\SN_\k$ (cfr. Lemma \ref{lemma-line}).
	\\
	To carry on the computation,  we consider the array $\PREC$ 
	such that for any 
	$i$	the value  $\PREC[i]$ represents the maximum index $j$ with  $\k\leq j\leq i$ such that 
	$\SN(j)=\SN-1$ (if none exists, we define $\PREC[i]=\k$).
	This value is used to check if any of the values  needed to compute $\min_j \SN(j)$ in (\ref{eq-Si}) is 
	$\SN_\k-1$, that is, the smaller of the two possibilities according to Lemma  \ref{lemma-line}.

	{\small
\begin{center}
\begin{tabular}{l}
\hline\\
{\bf Algorithm:}  LINE($L^n$) {\em [ Input: A path $L^n$ and the   array $D$ satisfying} (\ref{eq-D}){\em ]}\\

 {\bf \hphantom{1}1.} $\PREC[n-1]=n-1$, \   $\SN(n-1)=1$  $\ \ $ {\em [Here $d$ is the last node $n-1$]}\\
\\
 {\bf \hphantom{1}2.} $\k=\min \{i \ |\ D[i]=n-1\}$ $\qquad\ \ $ {\em [Now $d$ is the penultimate node of threshold 2 }\\
{\bf \hphantom{1}3.} $\SN(d)=2$, \ $\SN=2$   $\qquad\qquad\qquad\ \ $ {\em and we can set  $\SN(i)$ to 2  for each $i$ up to $n-2$;}\\
{\bf \hphantom{1}4.} {\bf for } $i=d+1$    {\bf to}  $D[d]-1$  $\qquad\ $ {\em moreover, we set   $\PREC[i]=i$  since all $\SN(i)$ have}\\
{\bf \hphantom{1}5.} \hphantom{F F F F}
 	$\SN(i)=2$,  \ $\PREC[i]=i$ $\qquad\ $ {\em  the same value $s$]}\\
	\\
{\bf \hphantom{1}6.} {\bf while }   $d>0$  {\bf do} \\
{\bf \hphantom{1}7.} \hphantom{F F F F}   $\k'=\k$\\
{\bf \hphantom{1}8.} \hphantom{F F F F}   $\k=\min \{i \ |\ D[i]=\k'\}$\\
{\bf \hphantom{1}9.} \hphantom{F F F F}   {\bf for } $i=d$    {\bf to}  $D[d]-1$  \\
{\bf 10.} \hphantom{F F   F F F F}   
								   {\em [We compute $\SN(i)\in\{\SN,\SN+1\}$ as in (\ref{eq-Si1}). Recall that here $D[i]=D[\k]$]}\\
{\bf 11.} \hphantom{F F   F F F F}
			   {\bf If}  $\SN(D[i])=\SN-1$  {\bf then } $\SN(i)=\SN$\\
{\bf 12.} \hphantom{F F F  F F F  F}
           {\bf else  } \\
{\bf 13.} \hphantom{F F F   F F F  F  E }
          Set  $j_{\min}=\max\big\{D[i]+1, \, 2D[i]-i-\lambda+1  \big\}$\\
{\bf 14.} \hphantom{F F  F  F F F  F  E }
			   Set   $j_{\max}=\min\big \{2D[i]-i+\lambda-1, \, D[D[i]]\big\}$\\
{\bf 15.} \hphantom{F F F  F F F  F E }
           {\bf If}  ($\SN(j_{\max})=\SN-1$ OR  $j_{\min}\leq \PREC[j_{\max}]$) {\bf then} $\SN(i)=\SN$ {\bf else}   $\SN(i)=\SN+1$\\
{\bf 16.} \hphantom{F F F F}  
						$\SN=\max\{ \SN(i) \ | \ d\leq i\leq D[d]\}$\\
{\bf 17.} \hphantom{F F F F}
        {\em [We set the values of the vector $L$ in the range $i= d, \ldots, D[d]$.}\\
{\bf 18.} \hphantom{F F F F}
        {\em $\ $Note that the values in the initial part of the interval are all set to $d$ as long as $\SN(i)=\SN$ ]}\\
{\bf 19.} \hphantom{F F F F}
         $\PREC[d]=d$\\
{\bf 20.} \hphantom{F F F F}
      {\bf For}   $i= d+1$ {\bf to }$D[d]$\\
{\bf 21.} \hphantom{F F F F F F  F}
             {\bf If}  $\SN(i)=\SN-1$  {\bf then } $\PREC[i]=i$ {\bf else} $\PREC[i]=\PREC[i-1]$.	\\	
        \hline
\end{tabular}	
\end{center}	
	}

\medskip
 \noindent
The algorithm takes time $O(1)$ to compute each $\SN(i)$,  
  $i=n-1,\ldots, 0$, for a total of $O(n)$. 
 \\ 
The actual  TWC target set of 
optimal size $\SN(0)$ can be constructed using standard backtracking techniques. \hfill $\Box$

{
\begin{example}\label{ex-linea}
Consider the path  of Figure \ref{fig3}(a). The threshold of each node 
is indicated in the circle inside the node itself.
Pruning the extremal nodes of threshold 1,  we get  a path of 20 nodes. 
Naming 
 them  from 0 to 19, we get the path  $L^{20}$ in Figure \ref{fig3}(b) where  
the  nodes of threshold 2 are $\k_1=0,\ \k_2=1,\ \k_3=5,\ \k_4=14$, and $\k_5=19$; hence, 
\\
$D[18]=D[17]=\ldots=D[14]=19, \ \ D[13]=\ldots =D[5]=14, \ \ D[4]=\ldots =D[1]=5,\ \ D[0]=1.$

\begin{figure}[ht!]\label{fig3}

\bigskip
\cnodeput[framesep=2pt](0,0){1}{\tiny $1$}
\cnodeput[framesep=2pt](.6,0){2}{\tiny $2$}
\cnodeput[framesep=2pt](1.2,0){3}{\tiny $2$}
\cnodeput[framesep=2pt](1.8,0){4}{\tiny $1$}
\cnodeput[framesep=2pt](2.4,0){5}{\tiny $1$}
\cnodeput[framesep=2pt](3,0){6}{\tiny $1$}
\cnodeput[framesep=2pt](3.6,0){7}{\tiny $2$}
\cnodeput[framesep=2pt](4.2,0){8}{\tiny $1$}
\cnodeput[framesep=2pt](4.8,0){9}{\tiny $1$}
\cnodeput[framesep=2pt](5.4,0){10}{\tiny $1$}
\cnodeput[framesep=2pt](6,0){11}{\tiny $1$}
\cnodeput[framesep=2pt](6.6,0){12}{\tiny $1$}
\cnodeput[framesep=2pt](7.2,0){13}{\tiny $1$}
\cnodeput[framesep=2pt](7.8,0){14}{\tiny $1$}
\cnodeput[framesep=2pt](8.4,0){15}{\tiny $1$}
\cnodeput[framesep=2pt](9,0){16}{\tiny $2$}
\cnodeput[framesep=2pt](9.6,0){17}{\tiny $1$}
\cnodeput[framesep=2pt](10.2,0){18}{\tiny $1$}
\cnodeput[framesep=2pt](10.8,0){19}{\tiny $1$}
\cnodeput[framesep=2pt](11.4,0){20}{\tiny $1$}
\cnodeput[framesep=2pt](12,0){21}{\tiny $2$}
\cnodeput[framesep=2pt](12.6,0){22}{\tiny $1$}
\cnodeput[framesep=2pt](13.2,0){23}{\tiny $1$}
\ncline{1}{2}
\ncline{2}{3}
\ncline{3}{4}
\ncline{4}{5}
\ncline{5}{6}
\ncline{6}{7}
\ncline{7}{8}
\ncline{8}{9}
\ncline{9}{10}
\ncline{10}{11}
\ncline{11}{12}
\ncline{12}{13}
\ncline{13}{14}
\ncline{14}{15}
\ncline{15}{16}
\ncline{16}{17}
\ncline{17}{18}
\ncline{18}{19}
\ncline{19}{20}
\ncline{20}{21}
\ncline{21}{22}
\ncline{22}{23}
\rput(6.6,-.6){(a)}
\cnodeput[framesep=2pt](.6,-1.5){2b}{\tiny $2$}
\cnodeput[framesep=2pt](1.2,-1.5){3b}{\tiny $2$}
\cnodeput[framesep=2pt](1.8,-1.5){4b}{\tiny $1$}
\cnodeput[framesep=2pt](2.4,-1.5){5b}{\tiny $1$}
\cnodeput[framesep=2pt](3,-1.5){6b}{\tiny $1$}
\cnodeput[framesep=2pt](3.6,-1.5){7b}{\tiny $2$}
\cnodeput[framesep=2pt](4.2,-1.5){8b}{\tiny $1$}
\cnodeput[framesep=2pt](4.8,-1.5){9b}{\tiny $1$}
\cnodeput[framesep=2pt](5.4,-1.5){10b}{\tiny $1$}
\cnodeput[framesep=2pt](6,-1.5){11b}{\tiny $1$}
\cnodeput[framesep=2pt](6.6,-1.5){12b}{\tiny $1$}
\cnodeput[framesep=2pt](7.2,-1.5){13b}{\tiny $1$}
\cnodeput[framesep=2pt](7.8,-1.5){14b}{\tiny $1$}
\cnodeput[framesep=2pt](8.4,-1.5){15b}{\tiny $1$}
\cnodeput[framesep=2pt](9,-1.5){16b}{\tiny $2$}
\cnodeput[framesep=2pt](9.6,-1.5){17b}{\tiny $1$}
\cnodeput[framesep=2pt](10.2,-1.5){18b}{\tiny $1$}
\cnodeput[framesep=2pt](10.8,-1.5){19b}{\tiny $1$}
\cnodeput[framesep=2pt](11.4,-1.5){20b}{\tiny $1$}
\cnodeput[framesep=2pt](12,-1.5){21b}{\tiny $2$}
\ncline{2b}{3b}
\ncline{3b}{4b}
\ncline{4b}{5b}
\ncline{5b}{6b}
\ncline{6b}{7b}
\ncline{7b}{8b}
\ncline{8b}{9b}
\ncline{9b}{10b}
\ncline{10b}{11b}
\ncline{11b}{12b}
\ncline{12b}{13b}
\ncline{13b}{14b}
\ncline{14b}{15b}
\ncline{15b}{16b}
\ncline{16b}{17b}
\ncline{17b}{18b}
\ncline{18b}{19b}
\ncline{19b}{20b}
\ncline{20b}{21b}
\rput(.6,-2.0){\tiny $0$}
\rput(1.2,-2.0){\tiny $1$}
\rput(1.8,-2){\tiny $2$}
\rput(2.4,-2){\tiny $3$}
\rput(3,-2){\tiny $4$}
\rput(3.6,-2){\tiny $5$}
\rput(4.2,-2){\tiny $6$}
\rput(4.8,-2){\tiny $7$}
\rput(5.4,-2){\tiny $8$}
\rput(6,-2){\tiny $9$}
\rput(6.6,-2){\tiny $10$}
\rput(7.2,-2){\tiny $11$}
\rput(7.8,-2){\tiny $12$}
\rput(8.4,-2){\tiny $13$}
\rput(9,-2){\tiny $14$}
\rput(9.6,-2){\tiny $15$}
\rput(10.2,-2){\tiny $16$}
\rput(10.8,-2){\tiny $17$}
\rput(11.4,-2){\tiny $18$}
\rput(12,-2){\tiny $19$}
\rput(6.6,-2.5){(b)}

\vspace*{2.7truecm}

\caption{}

\end{figure}

\medskip
\noindent
Let the \span size be $\lambda=2$.
Denote by $\SN(i)$ the cardinalities of the optimal 
solutions to the subproblems 
and by $\SS_i$ a corresponding optimal set (recall that  in 
general  $\SS_i$ is not unique).
We have

\begin{center}
\begin{tabular}{lll}
  $\SN(19) = 1$, & \mbox{and} & $\SS_{19}=\{19\},$ \\
  $\SN(i) = 1+\SN(19)= 2,$ & \mbox{and} &  $\SS_i=\{i,19\}, \ \mbox{ for }  i=18,17,\ldots,14,$ \\ 
  $\SN(i) = 1+\SN(14)= 3,$ & \mbox{and} &  $\SS_i=\{i,14, 19\}, \ \mbox{ for }  i=13,12,11,$ \\ 
  $\SN(i) = 1+\SN(19)= 2,$ & \mbox{and} &  $\SS_i=\{i,19\}, \ \mbox{ for }  i=10,9,8,$ \\ 
  $\SN(i) = 1+\SN(14)= 3,$ & \mbox{and} &  $\SS_i=\{i,14, 19\}, \ \mbox{ for }  i=7,6,5,$ \\ 
   $\SN(4) = 1+\SN(5)= 4,$ & \mbox{and} &  $\SS_4=\{4,5,14, 19\}, $ \\ 
   $ \SN(3) = 1+\SN(8)= 3,$ & \mbox{and} &  $\SS_3=\{3,8, 19\}, $ \\ 
   $ \SN(2) = 1+\SN(8)= 3,$ & \mbox{and} &  $\SS_2=\{2,8, 19\}, $ \\ 
   $ \SN(1) = 1+\SN(9)= 3,$ & \mbox{and} & $\SS_1=\{1,9, 19\}, $ \\ 
   $ \SN(0) = 1+\SN(3)= 4,$ & \mbox{and} &  $\SS_0=\{0,3,8, 19\}$.
   \end{tabular}
    \end{center}
  Hence,  the  optimal  size of a TWC target set for the considered line is $\SN(0)=4$
  and an
  optimal set is $\SS_0=\{0,3,8, 19\}$.
  
 \end{example}
 }

\subsubsection{Rings}

 We can use Theorem \ref{teo-line}  above to design
 an algorithm for the \PROB problem on  rings. Let $R^n$ denote the ring on $n$
nodes $\{0,\ldots,n-1\}$ with edges $(i,(i+1) \bmod n)$ and thresholds $t(i)$,
for $i=0,\ldots, n-1$.

We first notice that if all nodes have threshold 2, then an optimal TWC target set
for $R^n$ trivially has size $\lceil n/2\rceil$, so let us now assume that
there exists a node $j$ that  has threshold $t(j)=1$.
Either $j$ is in an optimal TWC target set for $R^n$ or it is not.
Consider the path  $R_{j,2}^n$ obtained by
``breaking'' the ring $R^n$ at  node $j$,  duplicating node $j$  into $j$ and $j'$, and assigning
threshold 2 to both $j$ and $j'$
(regardless of the original threshold value $t(j)=1$ in  $R^n$).
Therefore, the
edges of $R_{j,2}^n$ are $(j,j+1), (j+1, j+2), \ldots, (n-2,n-1), (n-1,0), \ldots (j-2, j-1), (j-1, j')$.
The  thresholds of $R_{j,2}^n$ are  

$$t_{j,2}(i)=\begin{cases}{t(i)} & {\mbox{if  $0\leq i\leq n-1$ and $i\neq j$}}\cr
                                                                    {2} &  {\mbox{if $i=j$ or $i=j'$}.}\end{cases}$$
 We can use the algorithm of Section \ref{sec:path} to compute the  size 
of an optimal TWC target set $\SS_{j,2}$ for the path $R_{j,2}^n$.
Notice that both $j$ and $j'$ must be in $\SS_{j,2}$, so $\SS_{j,2}-\{j'\}$ is  a 
TWC target set for the ring $R^n$,  optimal  among all TWC
target sets that include node $j$.

Now we want to  compute a TWC target set for the ring $R^n$ that is  optimal  among all TWC
target sets that do not include node $j$.
To do this, consider the  path  $R_{j,1}^n$ that has the same nodes and edges as $R_{j,2}^n$
but has  thresholds

$$t_{j,1}(i)=\begin{cases}{t(i)} & {\mbox{if  $0\leq i\leq n-1$ and $i\neq j$}}\cr
                                                                    {1} &  {\mbox{if $i=j$ or $i=j'$}.}\end{cases}$$
In particular, the endpoints of $R_{j,1}^n$ have thresholds   $t_{j,1}(j)=t_{j,1}(j')=1$.
First, we apply Lemma~\ref{claim-1} to $R_{j,1}^n$ and then we use the 
algorithm of  Section \ref{sec:path} to compute (the size of)
an optimal TWC target set $\SS_{j,1}$.
Since $j,j'\notin \SS_{j,1}$, we have that $\SS_{j,1}$ is  a 
TWC target set for the ring $R^n$,  optimal  among all TWC
target sets that do not include node $j$.

An optimal solution for the ring $R^n$ is then obtained by choosing the smaller
of $\SS_{j,2}-\{j'\}$ and $\SS_{j,1}$.
In conclusion we have the following result.
 \begin{theorem}\label{cor-ring}
 For any value of the time window size  $\lambda$, an optimal
 TWC target set for the ring $R^n$  can be computed in time $O(n)$.
   \end{theorem}

\subsection{Trees}

Let $T=(V,E)$ be a tree with threshold function $t: V \longrightarrow \N $,
 and let $\lambda\geq 1$ be a fixed  value of the time window size.
   We consider  $T$ to be rooted at some arbitrary 
 node $\root\in V$. For each  node $v\in V$, we denote by $T_v=(V_v,E_v)$ the 
 subtree of $T$ 
rooted at $v$.  
Moreover, we denote by $\Ch(v)$ the set of all  children of 
node $v$ in $T_v$.

\begin{definition}\label{def-S}
Given  node $v\in V$ and integers $t, r$, with  $t\in \{t(v), t(v)-1\}$ and $\i\geq 0$, 
we denote by 
 $\SN(v,t,\i)$ the minimum size of a TWC target set $\SS\subseteq V_v$ for   
 subtree $T_v$ that influences node $v$ in round 
$\i$ (that is,
       $v\in\influenced[\SS,\i] \setminus \influenced[\SS,\i-1]$),
         under the assumption  that  $v$ has  threshold  $t$  in $T_v$.
The threshold of each  other 
         node $w\neq v$ in $T_v$ is its original one $t(w)$.
\end{definition} 
The size of  an optimal TWC target set for  the tree $T$  can be  computed as 
\begin{equation}\label{eq-s-tree}
\min_{\i} \SN(\root,t(\root),\i),
\end{equation}
where $\i$ ranges between $0$ and the maximum possible number of 
rounds needed to complete the influence diffusion process.
The  number of rounds   is always upper bounded by the number of 
nodes in the graph (since 
 at least one new node must be influenced in each round before the 
 diffusion process stops). However, for  a tree $T$,
 this value is upper bounded by the length of the 
 longest path in $T$. In other words,
 the parameter $\i$ in Definition \ref{def-S} is bounded by the diameter
 $\diameter(T)$ of $T$.

We use a dynamic programming approach to compute the value in 
(\ref{eq-s-tree}). Then, the corresponding optimal TWC target set $\SS$ can be 
built using standard backtracking techniques. In our dynamic programming algorithm 
we compute all of the values 

$\qquad$$\SN(v,t,\i)$ for each $v\in V$, each $t\in \{t(v), t(v)-1\}$, and   
$\i=0,\ldots, \diameter(T)$, \\
and the  computation is performed according to a breadth-first  search 
(BFS) reverse ordering of the nodes of $T$, 
  so that each node $v$ is considered only when  all of
  the values  $\SN(\cdot,\cdot,\cdot)$ for all of its children are known. 
  The rationale behind the computation of both $\SN(v,t(v),\i)$ and 
  $\SN(v,t(v)-1, \i)$ is the following:
\\
i) $\SN(v,t(v),\i)$ corresponds to the case of a target set $\SS$ for 
 tree $T$  such that 

-- $v\in\influenced[\SS,\i]\setminus\influenced[\SS,\i-1]$
and 

-- at least $t(v)$ of $v$'s children belong to $\active[\SS\cap V_v,\i]\subseteq \active[\SS,\i]$;

\noindent
ii) $\SN(v,t(v)-1,\i)$ is the  size  of an optimal target set $\SS$ for $T$ satisfying 

-- $v\in\influenced[\SS,\i]\setminus \influenced[\SS,\i-1]$,

-- $\active[\SS,\i]$ contains $v$'s parent in $T$, and 

--  at least $t(v)-1$ of $v$'s children belong to 
$\active[\SS\cap V_v,\i]\subseteq \active[\SS,\i]$.

In the following, we show how to compute the above values $\SN(\cdot,\cdot,\cdot)$.
The procedure is summarized in algorithm TREE.

{\small
\begin{center}
\begin{tabular}{l}
\hline\\
{\bf Algorithm TREE($T,\root,\lambda, t$)}  
 {\em [{\tt Input:} Tree $T$ rooted at $\root$, 
\span 
size $\lambda$, threshold function $t$.]}\\
{\bf $\ $1.}
{\bf For} each $v\in T$ in reverse order to a BFS of $T$
\\
{\bf $\ $2.}
\hphantom{F } {\em [We  compute $\SN(v,t,\i)$ for each 
$t\in \{t(v), t(v)-1\}$ and $0\leq \i\leq\ \diameter(T)$]}
\\
{\bf $\ $3.}
\hphantom{F }              {\bf If}     $v$ is  a leaf {\bf then} {\em[ here $t(v)=1$]} \\
{\bf $\ $4.}
\hphantom{F  I  }   {\bf For}  $\i=0,\ldots, \diameter(T)$  \\
{\bf $\ $5.}
\hphantom{F  I FO }  Set  $\ \SN(v,0,\i)=0$  and $\SN(v,1,\i)=\begin{cases}{1}& { \hbox{if  } \i=0}\\ {\infty}&{\hbox{otherwise} }\end{cases}\ $ \\
{\bf $\ $6.}
\hphantom{F } {\bf If } $v$ is NOT a leaf in $T$ {\bf then} \\
{\bf $\ $7.}
\hphantom{F F} {\bf For } $\i=0,\ldots,  \diameter(T)$  AND $t\in \{t(v), t(v)-1\}$  (only $t=t(v)$ if $v=\root$)  \\ 
{\bf $\ $8.}
							\hphantom{F f f I } {\bf If} $\i=0$ OR $t\leq 1$ {\bf then}\\
{\bf $\ $9.}
							\hphantom{F f f I I  I } 	{\bf For} each  $w\in \Ch(v)$ compute\\
{\bf 10.} \hphantom{F f f I I I I I I }  
             { $min(w)=\min\left\{\min_{\i+1\leq j\leq \i+\lambda} \SN(w, t(w)-1,j), \min_{\i-1\leq j\leq diam(T)} \SN(w,t(w),j)\right\}$}\\
{\bf 11.}
\hphantom{F f f I I I } {\bf If} $\i=0$  {\bf then}   Set $\SN(v,t,0)=1+\sum_{w\in\Ch(v)} min(w)$  \\
{\bf 12.}
\hphantom{F f f I I I } {\bf If} $\i\geq 1$ and $t=0$ {\bf then}  Set $\SN(v,0,\i)=\sum_{w\in\Ch(v)} min(w)$  \\
{\bf 13.}
\hphantom{F f f I I I } 
                            		{\bf If} $\i\geq 1$ and $t= 1$ {\bf then}\\
{\bf 14.}
\hphantom{F f f I  I I IF }
											Set $z=\arg\!\min_{w\in \Ch(v)}\{ \SN(w,t(w), \i-1)-min(w)\}$\\
{\bf 15.}
\hphantom{F f f I I I IF }
											Set  $\SN(v,1,\i)=\sum_{w\in\Ch(v)\setminus \{z\}}m(w)+\SN(z,t(z),\i-1)$\\
 
{\bf 16.}
 \hphantom{F f f I } 
                            		{\bf If} $\i \geq 1$ and $t> 1$ {\bf then}\\
{\bf 17.}
\hphantom{F f f  I IF } 
 					Set $\SN(v,t,\i)=\min \sum_{w\in\Ch(v)} m(w)$, where\\
{\bf 18.}
\hphantom{F  f f I IF } 
				$m(w)\in \{ \SN(w, \tau,j)\  : \  ( \tau=t(w) \mbox{ AND } j\geq 0 ) \mbox{ OR } 
				                                          ( \tau=t(w)-1 \mbox{ AND } \i<j\leq \i+\lambda)  \}$\\                          
{\bf 19.}
\hphantom{F f f  I IF } 
         $|\{w\  :\  m(w)=\SN(w, t(w),j), \  \i-\lambda \leq j\leq \i-1\}|\geq t $ \\
{\bf 20.}
\hphantom{F  f f I IF }
        $|\{w\ :\ m(w)=\SN(w, t(w),j), \  \ell-\lambda \leq j\leq \ell-1\}|<  t$, $\quad\forall \ell=1,\ldots, \i-1$\\
        \hline
\end{tabular}	
\end{center}	
}

\smallskip

First, consider the computation of $\SN(v,t,\i)$ when   $v$ is a 
leaf of  $T$ in lines {\bf 3.} to {\bf 5.} In this case we have  $t(v)=deg(v)=1$. \\
-- If $\i=0$,   $v$  trivially must   belong to the target set
since  $v$ needs to be active at time 0; hence $\SN(v,t,0)=1$. 
\\
-- If $\i>0$ and $t=t(v)=1$, we observe  that 
any TWC target set that  influences leaf  $v$ at 
time \emph{exactly} $\i$ cannot contain $v$ and,
therefore, must influence $v$'s parent at time $\i-1$.
To do so, we set  $\SN(v,1,\i)=\infty$
in the algorithm; this  
forces the minimum at line {\bf 10.} or {\bf 17.}  to be reached with threshold 
$t(v)-1=0$, thus forcing
$v$'s parent to be active in round $\i-1$.\\
-- If $\i>0$ and $t=t(v)-1=0$ then, trivially, $\SN(v,0,\i)=0$.

\smallskip

Now consider an arbitrary  internal node $v$. Since we process nodes in a 
BFS reverse order, each child of $v$ has already been processed
 when the algorithm processes $v$.  
If $\i=0$, then $v$ must  necessarily  be in the target set and any 
$w\in\Ch(v)$  can benefit  from this. Therefore, the 
size $\SN(v,t,0)$ of an optimal solution for the subtree $T_v$ (line {\bf 11.})
for both $t=t(v)$ and $t=t(v)-1$ is equal to 
 $$\SN(v,t,0)=1+\sum_{w\in\Ch(v)} min(w)=1+\sum_{w\in\Ch(v)} \min\left\{\min_{1\leq j\leq \lambda} 
 \SN(w, t(w)-1,j), \min_{0 \leq j \leq \diameter(T)} \SN(w,t(w),j)\right\}.$$
Notice that we have constrained $j$ to be in the range $1, \ldots , \lambda$
in the formula above when $w$'s threshold is $t(w)-1$.
 This is correct since $v$ is active and able to  influence  $w$
only in rounds $j=1,\ldots, \lambda$.

Now, let us consider the computation of $\SN(v,t,\i)$ with  $\i\geq 1$, that is,
when $v$  
is not part of the target set and $v$
is influenced at time $\i$ by $t$ of its children 
(plus its parent if $t=t(v)-1$).
To determine the optimal solution, we need to know
the best  among the values  $\SN(w,\tau,j)$ for each $w\in\Ch(v)$ and  
 for all possible 
values of parameters $\tau$ and $j$, 
subject to the following two constraints:
\begin{itemize}
\item[\textbf{1)}] if $\tau=t(w)-1$, then $\i+1\leq j \leq \i+\lambda$
  (indeed $v$ is active and can influence  $w$ only during the $\lambda $ 
  rounds after it has become influenced, that is, in rounds $j=\i+1,\ldots,  \i+\lambda$),  
\item[\textbf{2)}]
at least  $t$ nodes  in $\Ch(v)$ are active in round $\i$ but at
 most $t-1$ are active in
any previous round $j\leq \i-1$
(otherwise $v$ would become influenced before the required round $r$).
\\
The special case $t=0$ (line  {\bf 12.}) can hold only if $t(v)=1$ and $t=t(v)-1=0$; 
hence  node $v$ must be  influenced by its parent at round $\i$ 
and none of its children can be active before round $r$.
Recall that $\SN(v,t,\i)$ asks for the minimum size of a target set  in $T_v$ (or equivalently a target set for $T$ 
having the smallest possible intersection  with $V_v$)  which makes $v$ have at least $t$ active children  at  exactly round $\i$.\\
\end{itemize}

\noindent
More formally, in the algorithm  we  compute  

\begin{eqnarray}
&&\SN(v,t,r)= \min  \sum_{w\in \Ch(v)} m(w)  \label{minS}\\
\nonumber && \mbox{where the following three conditions must be satisfied} \\
 \nonumber  && m(w)\in \{ \SN(w, \tau,j)\  : \  ( \tau=t(w) \mbox{ AND } j\geq 0 ) \mbox{ OR } 
				                                          ( \tau=t(w)-1 \mbox{ AND } \i+1\leq j\leq \i+\lambda)  \},\\     
 \nonumber && |\{w\  :\  m(w)=\SN(w, t(w),j), \  \max\left\{0,\i-\lambda\right\} \leq j\leq \i-1\}|\geq t,  \\
   \nonumber  &&  |\{w\ :\ m(w)=\SN(w, t(w),j), \  \max\left\{0,\ell-\lambda\right\} \leq j\leq \max\left\{0,\ell-1\right\}\}|<  t, 
\forall \ell=1,\ldots, \i-1.
\end{eqnarray}
\remove{\vspace{-0.5truecm}
\begin{example}\label{ex-4}
Consider the tree $T$ given in Figure ??? of Example \ref{ex-1a}. Recall that $\lambda =1$.
We have that  ....
\end{example}
%
}

\begin{lemma}\label{lemmaLP}
The value in (\ref{minS}) can be computed in polynomial time.
\end{lemma}

\proof
Consider the computation of $\SN(v,t,\i)$. Define the variables
 $x(w,t,j)$, for each  $w\in\Ch(v)$ and each choice of 
$t\in\{t(w)-1,t(w)\}$ and $j\geq 0$ such that
if $t=t(w)-1$ then  $\i+1\leq j\leq \i+\lambda$.
We can re--write (\ref{minS}) as
\begin{eqnarray*}
&&\mbox{\bf Compute}\\
&&\min  \sum_{w\in\Ch(v)} \left(\sum_{j\geq 0}x(w,t(w),j)\SN(w,t(w), j)                               +\sum_{j=\i+1}^{\i+\lambda}x(w,t(w)-1,j)\SN(w,t(w)-1,j)\right)\\
&&\mbox{\bf subject to}\\ 
&& x(w,t,j) \in\{0,1\},   \qquad w\in\Ch(v),\ t=t(w),t(w)-1,\ j=0,\ldots,\diameter(T)\\
&& \sum_{j=0}^{\diameter(T)}x(w,t(w),j) + \sum_{j=\i+1}^{\i+\lambda} x(w,t(w)-1,j)=1,                       \qquad  \mbox{for each } w\in\Ch(v)\\ 
 && \sum_{w\in \Ch(v)}\sum_{j=\i-\lambda}^{\i-1}x(w,t(w),j) \geq t\\
 && \sum_{w\in \Ch(v)}\sum_{j=\ell-\lambda}^{\ell-1}- x(w,t(w),j) \geq -(t-1),   \qquad \mbox{for each }\ell=\lambda, \ldots, \i-1\\
\end{eqnarray*}
The coefficient matrix $M$ of the above linear program is shown in Figure \ref{fig4}.
 Matrix $M$  satisfies the Ghouila-Houri characterization of  totally 
 unimodular matrices:
 each subset $R$ of rows of $M$ can be partitioned into   $R_1$ and $R_2$ 
 so that all the entries of 
 the  vector $\sum_{\r \in R_1} \r- \sum_{\r \in R_2} \r$  belong to  
 $\{-1,0,1\}$ \cite{Schrijver-03}.
 (As a matter of fact, one can show that the above linear
program corresponds to a minimum cost flow problem in an
associated auxiliary network.) Therefore, the linear program can be optimally solved  
in polynomial time~\cite{Schrijver-03}. 

\noindent
\begin{figure}[ht!]\label{fig4}
\noindent
{\scriptsize{
$\bordermatrix{  
  &\ldots &w(0)&w(1) &\ldots &w(\lambda -1) &w(\lambda )&\ldots &w(\i-\lambda-1)&w(\i-\lambda)&\ldots &w( \i-2)&w( \i-1)&{{ w(j) \atop{\ldots \qquad\quad \ldots\atop {  j\geq \i}}} }   & {  w'(j) \atop{\ldots \qquad\qquad \ldots\atop{\i\leq j\leq \i+\lambda}}}      \cr
\cr
&\ldots &1   &1  &\ldots & 1 & 1              & \ldots  &1&1&\ldots &1&1& 1&  1 \cr
\cr
%
&\ldots&0     &0 & \ldots &0 & 0& \ldots & 0& 1&\ldots &1& 1& 0&  0 \cr
\cr
&\ldots&-1 & -1 &\ldots & -1 & 0              & \ldots &0&0&\ldots &0&0& 0& 0 \cr
&\ldots&0 & -1 &\ldots & -1 & -1              & \ldots &0&0&\ldots &0&0& 0& 0 &\cr
&\ddots&\vdots &\vdots &\ddots &\vdots   &\vdots&\ddots    &\vdots &\vdots&\ddots&\vdots&\vdots&  \vdots&  \vdots\cr
&\ldots&0    & 0 &\ldots &0 & 0           & \ldots &-1&-1&\ldots & -1&0& 0& 0 \cr
\cr		
}$
}}
\caption{The portion of the  matrix $M$ made of columns corresponding to variables $x(w,t,j)$ 
                     for a fixed node $w$: here $w(j)=x(w,t(w),j)$ and $w'(j)=x(w,t(w)-1,j)$. $\quad\quad$}
\end{figure}

\noindent
The algorithm TREE($T,\root,\lambda$) requires $O(n^2)$ computations of values 
$\SN(v,t,r)$ using (\ref{minS}).
Hence we get the following result.

\begin{theorem}
For any tree $T$, the optimal TWC target set  can be computed in polynomial time.
\end{theorem}

\subsection{Complete Graphs}

Let $K_n=(V,E)$ denote the complete graph on $n$ nodes.
The following observation was  made in \cite{NNUW} for 
target set selection without a time window constraint;
it is easy to  see that it also holds in our scenario.

\begin{lemma}\label{claim-2}{\rm \cite{NNUW}}
If the optimal TWC target set for $K_n$ has size $k$, then there exists an optimal TWC 
target set consisting 
of $k$ nodes with the largest thresholds in $K_n$.
\end{lemma}

Lemma~\ref{claim-2} follows from the observation that 
in   any target set $\SS$ for $K_n$, 
if there exist $v\in \SS$ and $u\in V-\SS$ with $t(v)<t(u)$, then
 $\SS\setminus\{v\}\cup\{u\}$ is also a target set for $K_n$.
Lemma \ref{claim-2} implies that we only  need to determine the size 
of an optimal TWC target set.
Let  $A[1\ldots n-1]$ denote the integer vector such that 
$A[\ell]=|\{v\in V\ |\ t(v)\leq \ell\}|$, for $\ell=1,\ldots,n-1$.
As in counting sort, the vector $A[1\ldots n-1]$ can be computed in linear time.
The following algorithm $MAX(n,k)$  determines the largest 
number of nodes that can be influenced
using a TWC target set of  $k$ nodes. 

\medskip
\begin{tabular}{l}
\hline
{\bf Algorithm $MAX(n,k)$} 
 \small{{\em [Input:  $n$, vector $A[1\ldots n-1]$, 
  \span size  $\lambda$, and size $k$]}}\\
{\bf $\ $1.}
Set $\ell=k$\\
{\bf $\ $2.}
{\bf If} $A[\ell]>0$ {\bf then} {\em [at least one node outside the target set  can be influenced]} \\
{\bf $\ $3.}
\hphantom{F    }   {\bf For}  $j=0,\ldots, \lambda -2$ \\
{\bf $\ $4.}
\hphantom{F   FO }  Set  $X[j]=-k$ \\
{\bf $\ $5.}
\hphantom{F    } Set $X[\lambda -1]=0$\\
{\bf $\ $6.}
\hphantom{F    } Set $j=0$;  \\
{\bf $\ $7.}
\hphantom{F }
               {\bf Repeat} \\
{\bf $\ $8.}
\hphantom{F  R  }  Set $y=A[\ell]$\\
{\bf $\ $9.}
\hphantom{F  R  }  Set $\ell=A[\ell]-X[j]$  \\
{\bf $\ $10.}
\hphantom{F   }  \, Set $X[j]=y$  \\
{\bf $\ $11.}
\hphantom{F   } \, Set $j=(j+1) \bmod \lambda$  \\
{\bf $\ $12.}
\hphantom{I } 	{\bf Until} ($A[\ell]-X[j]\leq \ell$ OR $A[\ell]+k\geq n$)\\
{\bf $\ $13.}  
 {\bf Output }  $\min\{n, k+A[\ell]\}$
 \\
 \hline
\end{tabular}	

\begin{lemma}
For any given integer  $k$, the algorithm $MAX(n,k)$ computes  the 
largest number of nodes that can be influenced in $K_n$
using a TWC target set of size $k$.
\end{lemma}
\proof
Let $\SS$ denote a set of $k$ nodes with the largest thresholds. Let $\i\geq 0$
and consider the algorithm $MAX(n,k)$.
By induction on the number   $\i$ of  iterations of 
the \textbf{repeat} cycle one can see that:
\begin{itemize}
\item The value of the variable  $\ell$ is the size of  $\active[\SS,\i]$, i.e., the number 
of active nodes in round $\i$ of the influence  process in $K_n$ with  TWC 
target set $\SS$.
\item If $\i\geq \lambda -1$, then  the value of $X[\i \bmod \lambda]$ equals 
the number  $|\influenced[\SS,\i-\lambda]|$ of nodes that have been 
influenced during round $\i- \lambda$. 
\item
The difference $A[\ell]-X[\i \bmod \lambda]$ is the number of 
nodes that can be influenced in round $\i+1$, that is,  
$|\influenced[\SS,\i+1]\setminus \influenced[\SS,\i]|$. \hfill $\Box$
\end{itemize}

\noindent
Algorithm $MAX(n,k)$ requires  $O(n)$ time; using a   binary 
search  for the optimal value of $k$ we obtain the following result. 
\begin{theorem}
The optimal TWC  target set in a complete graph $K_n$ can be computed 
in time $O(n\, \log n )$.
\end{theorem}

\section{Concluding Remarks}

In this paper we have introduced a new model of information diffusion 
in social networks 
in which agents change behaviors on the basis
of information collected from their neighbors in a time interval of bounded size,
unlike previous models in which agents have unbounded memory.
A number of interesting problems remain open.
First  of all, we would like to know whether our result on trees 
can be extended to graphs of bounded tree-width.
Secondly, it seems to us that a linear time algorithm for complete graphs
is possible, but it has so far escaped our attempts. 
Finally, it would be interesting to find sharp  upper and lower bounds
on the cardinality of a minimum TWC-TSS, in terms of $\lambda$ and easily
computable parameters of the graph. This goal seems to be ruled out for arbitrary  
graphs 
by our Theorem \ref{teo1}, but it could be achievable for interesting
classes of graphs (cf.\ the work for the classical TSS problem in~ \cite{ABW-10}).

\end{document}